\documentclass[superscriptaddress,amsmath, amssymb, amsfonts, twocolumn, footinbib]{revtex4-1}
\usepackage[german,american,english]{babel}
\usepackage{graphicx}
\usepackage{graphics}
\usepackage{dcolumn}
\usepackage{multirow}
\usepackage{soul}
\usepackage{bm}
\usepackage{latexsym}
\usepackage{amssymb}
\usepackage{amsmath}
\usepackage{amsfonts}
\usepackage{layout}
\usepackage{verbatim}
\usepackage{epsfig}
\usepackage{graphicx}
\usepackage{amsbsy}
\usepackage{lipsum}
\usepackage{cancel}
\usepackage[toc]{appendix}
\usepackage{xcolor}

\newcommand{\bea}{\begin{eqnarray*}}
	\newcommand{\eea}{\end{eqnarray*}}
\newcommand{\bne}{\begin{equation*}}
\newcommand{\ede}{\end{equation*}}
\newcommand{\ba}{\arraycolsep 0.3ex \begin{array}{rl}}
\newcommand{\ea}{\end{array}}
\newcommand{\dps}{\displaystyle}
\newcommand{\bnen}{\begin{equation}}
\newcommand{\eden}{\end{equation}}
\newcommand{\bean}{\begin{eqnarray}}
\newcommand{\eean}{\end{eqnarray}}
\newcommand{\bsen}{\begin{subequations}}
	\newcommand{\esen}{\end{subequations}}

\newcommand{\bna}{\begin{array}}
	\newcommand{\eda}{\end{array}}
\newcommand{\bnm}{\begin{enumerate}}
	\newcommand{\edm}{\end{enumerate}}

\newcommand {\pd} [2] {\frac{\partial #1}{\partial #2}}

\usepackage[utf8]{inputenc}

\begin{document}

\title{Dominance of extrinsic scattering mechanisms in the orbital Hall effect: graphene, transition metal dichalcogenides and topological antiferromagnets}
\author{Hong Liu}
\affiliation{School of Physics and Australian Research Council Centre of Excellence in Low-Energy Electronics Technologies, UNSW Node, The University of New South Wales, Sydney 2052, Australia}
\author{Dimitrie Culcer}
\affiliation{School of Physics and Australian Research Council Centre of Excellence in Low-Energy Electronics Technologies, UNSW Node, The University of New South Wales, Sydney 2052, Australia}
\begin{abstract}
The theory of the orbital Hall effect (OHE), a transverse flow of orbital angular momentum (OAM) in response to an electric field, has concentrated on intrinsic mechanisms. Here, using a quantum kinetic formulation, we determine the full OHE in the presence of short-range disorder using 2D massive Dirac fermions as a prototype. We find that, in doped systems, extrinsic effects associated with the Fermi surface (skew scattering and side jump) provide $\approx 95\%$ of the OHE. This suggests that, at experimentally relevant transport densities, the OHE is primarily extrinsic. 
\end{abstract}

\date{\today}

\maketitle

\textit{Introduction}. The electrical manipulation of magnetic degrees of freedom has been continuously investigated since \O{}ersted discovered the deflection of a compass needle by a current-carrying wire. Modern research has focused on the magneto-electric coupling provided by the spin-orbit interaction \cite{AHE-RevModPhys.82.1539,SHE-RevModPhys.87.1213,Edelstein-PhysRevResearch.3.013275,Roadmap-SOT-Review}, while an energetic recent effort has been geared towards the electrical operation of orbital degrees of freedom in systems without spin-orbit coupling \cite{OAM-Exp-Tobias,Without-SOC-PhysRevB.104.195114,Exp-OHE-Ti-Nat-2023-Hyun-Woo,OHE-Weak-SOC-npj, VOHE-TMD-PRB-Paul-2020,IOHE-Metal-PRB-2018-Hyun-Woo, Exp-OT-PRR-2020,Exp-OT-CommP-2021-Byong-Guk}. 
This effort centres on the realization that Bloch electrons, possess an orbital angular momentum (OAM) about their centre of mass \cite{Yafet-1963}, which is in part related to the Berry curvature \cite{Mingche-1995-PhysRevLett.75.1348,Rev-JOP-2008-Mingche,Mingche-1996-PhysRevB.53.7010, GaneshSundaram1999, Quanzization-orbits-EPJB}. The OAM affects semiclassical quantization \cite{Rev-JOP-2008-Mingche, GaneshSundaram1999, Quanzization-orbits-EPJB}, contributes to the magnetization in certain materials \cite{lahiri_PRB2021_nonlinear,OT-FM-PRB-2021-YoshiChika,OM-SciRep-2017-Yuriy}, affects the Zeeman splitting of Dirac materials \cite{PhysRevX.8.031023-Ensslin, PhysRevLett.123.026803-Ensslin&Ihn} and contributes to the non-linear magneto-resistance, valley-Hall effect \cite{lahiri_PRB2021_nonlinear, NLVHE_2023, azadeh_MR_PRB2023}, and anomalous Nernst effect \cite{Xiaodi-PhysRevLett.97.026603}.

In a time-reversal symmetric system a finite OAM density can be generated by an electric field via the orbital Edelstein effect \cite{OEE-SciR-2015-Shuichi,MM-PRL-2016-Ivo,OEE-NL-2018-Shuichi,OEE-NatComm-2019-Peter,Exp-OEE-PRL-2022-Jinbo} or by separating electrons with different OAMs on different sides of the sample using an electric current. This is the orbital Hall effect (OHE) \cite{Orbitronics-PRL-2005-Shoucheng,IOHE-PRB-2021-Giovanni,Exp-BC-NatPhy-2022-Eli,Exp-OHE-Ti-Nat-2023-Hyun-Woo,MOHE-FM-PRB-2022-Peter, ISOHE-PRL-2018-Hyun-Woo,IOHE-3Dmetal-PRL-2008-Inoue,IOHE-Metal-PRB-2018-Hyun-Woo,IOHE-XIV-PRB-2021-Hyun-Woo,OC-Rev-EL-2021-Yuriy}, which has received considerable attention recently \cite{Exp-OT-PRR-2020,Exp-OT-CommP-2021-Byong-Guk,OT-NatComm-2021-Kyung-Jin,OOS-Cvert-2020-PRL-Mathias,CIAM-PRR-2020-Yuriy,OT-OEE-NatComm-2018-Haibo,OT-FM-PRB-2021-YoshiChika,Exp-BC-NatPhy-2022-Eli}. The OHE has been proposed as the mechanism behind the valley Hall effect observed in graphene and transition metal dichalcogenides (TMDs) \cite{VHE-PRL-WangYao, IOHE-PRB-2021-Giovanni}. Injection of an orbital current into a ferromagnet generates an orbital torque on local magnetic moments \cite{Exp-OT-PRR-2020,Exp-OT-CommP-2021-Byong-Guk,OT-NatComm-2021-Kyung-Jin,OOS-Cvert-2020-PRL-Mathias,CIAM-PRR-2020-Yuriy,OT-OEE-NatComm-2018-Haibo,OT-FM-PRB-2021-YoshiChika,Exp-BC-NatPhy-2022-Eli}, and the OHE may likewise be responsible for the large spin Hall effect observed recently \cite{MOHE-FM-PRB-2022-Peter,ISOHE-PRL-2018-Hyun-Woo,IOHE-3Dmetal-PRL-2008-Inoue,IOHE-Metal-PRB-2018-Hyun-Woo,IOHE-XIV-PRB-2021-Hyun-Woo,OC-Rev-EL-2021-Yuriy}. Remarkably, all recent theoretical work has focused on intrinsic OHE mechanisms \cite{BiTMD-OHE-PRB-2022-Giovanni&Tatiana,OHE-BiTMD-PRL-2021-Tatiana,OHE-PRB-2022-Manchon,OH-phase-TMD-PRB-2020-Tatiana,IOHE-TMD-PRB-Satpathy-2020}, while neglecting extrinsic disorder contributions. This absence is puzzling since extrinsic scattering mechanisms such as skew scattering and side jump are known to contribute to the anomalous and spin-Hall effects at the same order in the disorder strength as the intrinsic contribution \cite{Inoue-SHE-PhysRevB.70.041303, Inoue-AHE-PhysRevLett.97.046604, AHE-Dirac-PRB-2007-Sinova, IOHE-3Dmetal-PRL-2008-Inoue, Interband-Coherence-PRB-2017-Dimi, SNandy2019, OrtixCarmine2021, DuZZ2021NatureComm, DuZZ2021}. It is natural to expect a substantial disorder contribution in the OHE. This has enormous implications for experiment: a signal may appear to be intrinsic, that is, independent of the disorder strength, yet our work shows it can still be almost entirely due to disorder.

\begin{figure}[tbp!]
\begin{center}
\includegraphics[trim=0cm 0cm 0cm 0cm, clip, width=0.95\columnwidth]{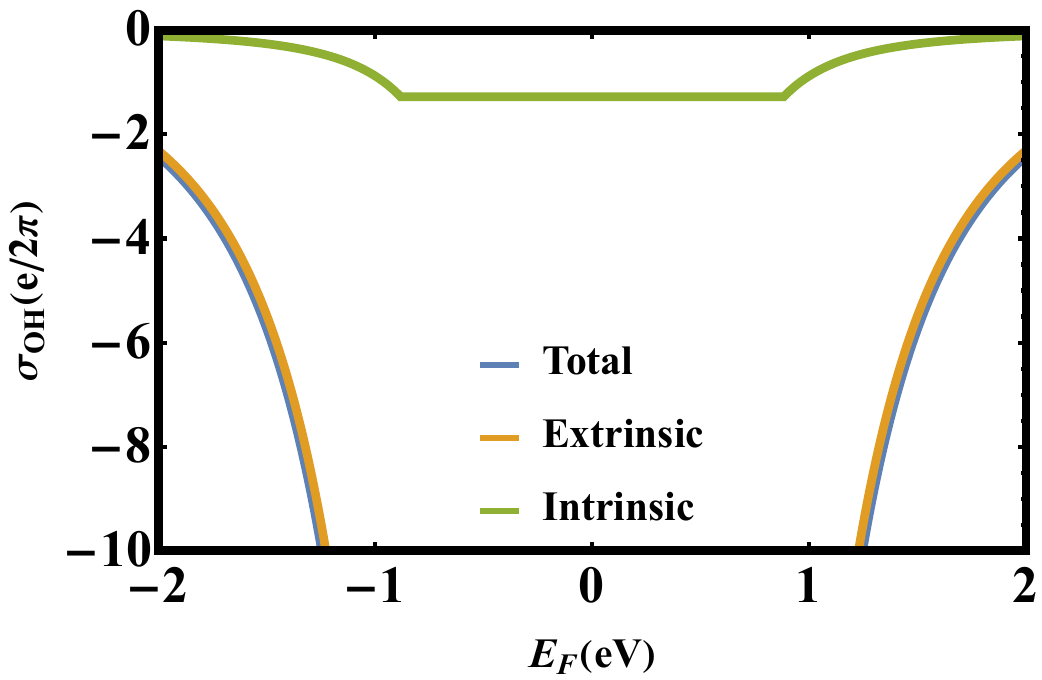}
\caption{\label{Fig1} The total, extrinsic and intrinsic OH-conductance $\sigma_\text{OH}$ vs the Fermi energy $E_F$ for the 2H-phase monolayer MoS$_2$. The parameters are Fermi velocity $v_F=3.6\text{eV}\AA$ and energy gap  $2\Delta=1.766\text{eV}$. 
}
\end{center}
\end{figure}

In this article, we calculate the full OHE, including intrinsic and extrinsic contributions, for two-dimensional (2D) massive  Dirac fermions, using graphene, TMDs, and topological antiferromagnets as prototype systems. We employ a quantum kinetic equation for the density matrix $\hat{\rho}$, derived from the quantum Liouville equation, following the blueprint of Refs.~\cite{Interband-Coherence-PRB-2017-Dimi, JE-PRR-Rhonald-2022}. 
Our main finding is that, remarkably, the extrinsic OH-conductance overwhelms the intrinsic OH-conductance in doped systems and provides the dominant contribution to the OHE when the Fermi energy lies in the conduction or valence bands. This finding overhauls the conventional interpretation of experimental data and suggests that what is measured experimentally is overwhelmingly the extrinsic contribution. For the model studied here of massive Dirac fermions, and assuming short-range impurities, the extrinsic contribution is 21 times of the intrinsic one. Our central result is summarized in Fig. ~\ref{Fig1} with the total, intrinsic and extrinsic OH-conductance $\sigma_\text{OH}$ vs Fermi energy plotted separately. The orbital Hall current operator $\hat{\bm j}=\frac{1}{2}\{\hat{\bm L},\hat{\bm v}\}$ where $\hat{\bm L}$ is the OAM operator and $\hat{\bm v}$ the velocity operator. With the applied electric field $E_x$ along the $\hat{\bm x}$-direction, the orbital Hall current $j^z_y$ for the $z$-component of the OAM flows along the $\hat{\bm y}$-direction, and $\sigma_\text{OH}=\sigma^z_{yx}=j^z_y/E_x$. The total OH-conductance
 \begin{equation}
 \ba
\sigma^\text{tot}_\text{OH}&\dps\approx-\frac{eg_sg_\nu}{12\pi} \frac{ m_e}{m^* }\frac{22\Delta^3}{(v^2_Fq^2_F+\Delta^2)^{3/2}},
\ea
 \end{equation}
where $m_e$ is the electron mass and $m^*=\hbar^2\Delta/v^2_F$ is the  effective mass of massive Dirac electron, $v_F$ is the Fermi velocity and $2\Delta$ is the energy gap and $g_s,g_\nu$ are spin/valley degeneracy. 
$q_F$ is defined by the Fermi energy $E_F=\sqrt{v^2_Fq^2_F+\Delta^2}$. The extrinsic contribution is Fermi surface effect $\sigma^\text{ext}_\text{OH}=\frac{21}{22}\sigma^\text{tot}_\text{OH}$. The intrinsic contribution is $\sigma^\text{int}_\text{OH}=\frac{1}{22}\sigma^\text{tot}_\text{OH}$. It is a continuous Fermi sea effect and attains the maximum absolute value when $E_F$ is within the energy gap $\sigma^\text{int}_\text{OH}=-\frac{eg_sg_\nu}{12\pi} \frac{m_e}{m^*}$. Physically, the dominance of extrinsic mechanisms reveals the strong role played by disorder-induced interband coherence in the transport of OAM. The OAM is an interband effect, and its intrinsic part is large in systems exhibiting large Berry curvatures, in other words, strong intrinsic interband coherence. On the other hand, skew scattering and side jump are sources of interband coherence mediated by disorder. Our calculation shows that, in the case of the OHE, this disorder-mediated contribution is one order of magnitude larger than the intrinsic terms, suggesting the best method to maximize the orbital Hall current may be to harness extrinsic mechanisms. Importantly, to obtain a large OHE it is \textit{not} necessary for the Berry curvature to be large, a finding that has profound implications for the electrical manipulation of magnetic moments. 
\begin{figure}[tbp!]
\begin{center}
\includegraphics[trim=0cm 0cm 0cm 0cm, clip, width=0.95\columnwidth]{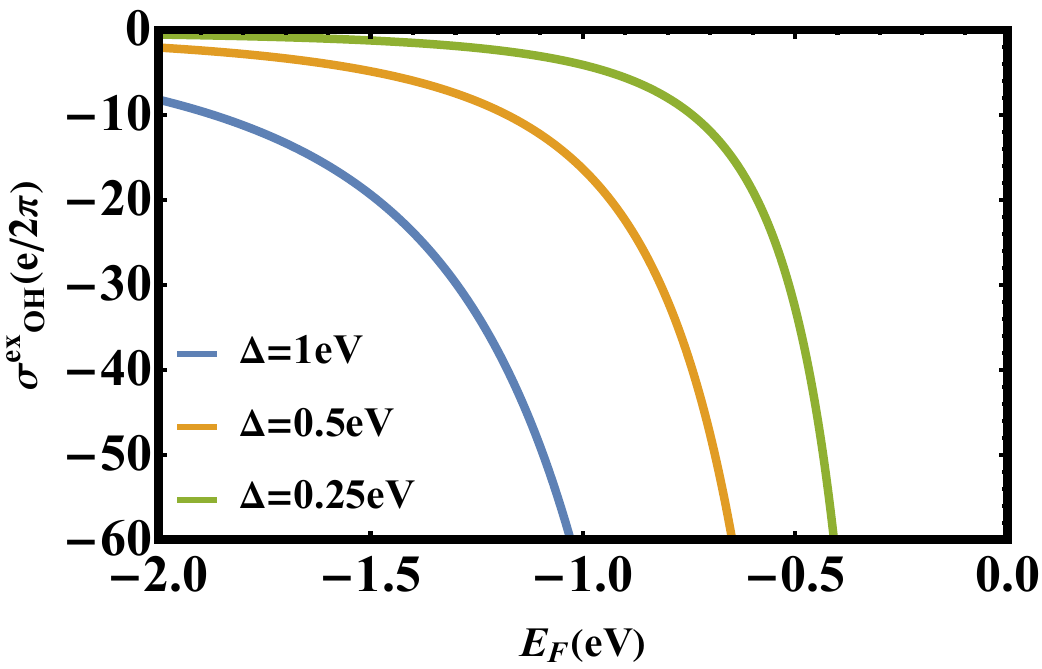}
\caption{\label{Fig2} The extrinsic OH-conductance $\sigma^\text{ext}_\text{OH}$ vs the Fermi energy $E_F$ for different energy band gaps. For graphene the parameters are $v_F=\frac{3}{2}at$ with $a=1.42\AA$ and $t=2.8\text{eV}$. 
}
\end{center}
\end{figure}

\textit{Kinetic equation}. We start with the quantum Liouville equation for the single-particle density operator $\hat{\rho}$,
\begin{equation}\label{QLE}
\pd{\hat{\rho}}{t}+\frac{i}{\hbar}\big[\hat{H},\hat{\rho}\big]=0,
\end{equation}
where the total Hamiltonian $\hat{H} = \hat{H}_0 + \hat{V} + \hat{U}$. $\hat{H}_0$ is the band Hamiltonian, $\hat{V}=e{\bm E}\cdot\hat{{\bm r}}$ is the external longitudinal electrical field, and $\hat{U}$ represents the disorder scattering potential, which we take to be scalar and short-range. We decompose the density matrix as $\hat{\rho}=\langle\hat{\rho}\rangle+\hat{g}_0$, where $\langle\hat{\rho}\rangle$ is averaged over disorder configurations, while $\hat{g}_0$ is the fluctuating part \cite{JE-PRR-Rhonald-2022}.
As shown in the Supplement, the disorder-averaged part $\langle\hat{\rho}\rangle$ satisfies
\begin{equation}
\pd{\langle\hat{\rho}\rangle}{t}+\frac{i}{\hbar}\big[\hat{H}_0,\langle\hat{\rho}\rangle\big]+J_0(\langle\hat{\rho}\rangle)=-\frac{i}{\hbar}\big[\hat{V},\langle\hat{\rho}\rangle\big]-J_E(\langle\hat{\rho}\rangle),
\end{equation}
where the electrical field corrected scattering term  $J_E(\langle\hat{\rho}\rangle)$ is given in the Supplement.

\textit{Graphene and TMDs.} $\hat{H}_0$ of monolayer graphene or TMDs has the form in momentum Pauli basis 
\begin{equation}
 H_0({\bm q})= v_F(\nu\tau_x q_x-\tau_y q_y) +\Delta\tau_z.
\end{equation}
$2\Delta$ represents a gap that can be induced by inversion/time reversal symmetry breaking, $\nu=\pm1$ is the valley index for materials with valley degree, $\tau_{x,y,z}$ are the Pauli matrices: in graphene or TMDs these represent the sub-lattice pseudospin, and $\arctan\phi=q_y/q_x$. The band dispersion $E^{m}_{\bm q}=m E_q $ with $m=\pm$ and $E_q=\sqrt{v^2_F q^2+\Delta^2}$. The symmetrized OAM operator is given by $\hat{\bm L}=\frac{m_0}{4}(\hat{\bm r}\times \hat{\bm v} - \hat{\bm v}\times\hat{\bm r})$ where $m_0=-\hbar/(g_L\mu_B)$ with g-factor $g_L=1$ and Bohr-magneton $\mu_B=\frac{e\hbar}{2m_e}$. The Berry connection is defined as $\bm{\mathcal{R}}^{mm'}_{\bm q}=i\langle u^m_{\bm q}|\nabla_{\bm q} u^{m'}_{\bm q}\rangle$ with $|u^m_{\bm q}\rangle$ the lattice periodic part of Bloch eigenstate wave function. The position operator is $ \langle u^m_{\bm q}|\hat{\bm r}|u^{m'}_{\bm q}\rangle =i  \pd{\delta(\bm{q}-\bm{q}^\prime)}{\bm{q}}\delta_{m,m'}+\bm{\mathcal{R}}^{mm'}_{\bm q}\delta(\bm{q}-\bm{q}')$ and the velocity operator is $\langle u^m_{\bm q}|\hat{\bm v}|u^{m'}_{\bm q}\rangle =\frac{1}{\hbar}\pd{E^m_{\bm q}}{q}\delta_{m,m'}+\frac{i}{\hbar}\big(E^m_{\bm{q}}-  E^{m'}_{\bm{q}}\big)\bm{\mathcal{R}}^{mm'}_{\bm q}$. Only the $z$-component of OAM for 2D massive Dirac model survives, $L_z=-(1/g_L\mu_B)(v^2_F\Delta/2 E^2_q)\tau_0$
\footnote{  The limit of $\Delta\rightarrow 0$ is related to the fact that the integral of a limit and the limit of an integral are generally not the same. This behaviour is also present in the Berry curvature, as discussed in Ref.~\cite{PhysRevB.68.045327} for anomalous Hall effect. In this paper, if we take the $\Delta\rightarrow 0$ at the very beginning of graphene Hamiltonian, the OHE disappears altogether.But if we take the $\Delta\rightarrow 0$ limit for the orbital moment, it yields $\text{lim}_{\Delta\rightarrow 0}\Delta/(q^2v^2_F+\Delta^2)=\pi\delta(q^2v^2_F)$. For the intrinsic case, the Fermi energy lies exactly at the Dirac point. The $\Delta\rightarrow 0$ limit will make the intrinsic OH-conductance diverge after performing the integral over the Fermi sea \cite{IOHE-PRB-2021-Giovanni}. Hence, formally, the intrinsic OH-conductance will diverge when the $E_F$ is in band gap and the extrinsic OH-conductance will vanish for graphene. However, in realistic samples, we expect that once the gap reaches below a certain value it will be overwhelmed by disorder and such a divergence will not be observable in practice. The limit of vanishing gap makes sense only for graphene which is a gapless material in a pristine situation. The $\Delta\rightarrow 0$ limit is not applicable to TMDs, since the band gap of monolayer TMDs is inherent to the band structure of materials \cite{BiTMD-OHE-PRB-2022-Giovanni&Tatiana}.}.

The disorder averaged density matrix $\langle\hat{\rho}\rangle=\rho_0+\rho_E$ with  $\rho_0$ the density matrix in equilibrium and correction by external electrical field $\rho_E=n_E+S_E$. The kinetic equation is solved in terms of band-diagonal contribution $n_E$ and band off-diagonal contribution $S_E$. The expectation value of the orbital Hall current is calculated as ${\bm j}=\text{Tr}[\langle\hat{\rho}\rangle\hat{\bm j}]$ and Tr represents the full operator trace, and we focus on $\hat{j}^z_y=1/2 \{\hat{L}_z,\hat{v}_y\}$. Transport theory requires $E_F \tau/\hbar \gg 1$, where $\tau$ represents a characteristic momentum relaxation time that can be used as a measure of the disorder strength. The theory is formulated as an expansion in the small parameter $\hbar/(E_F \tau)$. The first term leads to the Drude conductivity $\propto \tau$, formally of order $(-1)$ in the small parameter. The next term is of order zero in the disorder strength. Therefore, one finds $\rho_E=n^{(-1)}_E + S^{(0)}_E+ n^{(0)}_E$ up to zero order in the impurity density. The leading diagonal $n^{(-1)}_E$ representing the shift in the Fermi surface induced by the electric field is found from the band diagonal scattering integral $J^\text{1st}_0 [ n^{(-1)}_{{E}}]$. Next one feeds $n^{(-1)}_E$ into the band off-diagonal scattering integral $J^\text{1st}_0 [ n^{(-1)}_{{E}}]$ to determine the anomalous driving term $D'_{E}$, and ultimately the band off-diagonal density matrix $S^{(0)}_E$. At last, we feed $S^{(0)}_E$ into $J^{\text{1st}}_\text{dia}[S^{(0)}_E]$ to get the zero-order $n^{(0)}_E$. This contribution is contained in vertex corrections in the diagrammatic formalism ~\cite{Interband-Coherence-PRB-2017-Dimi, JE-PRR-Rhonald-2022}. These corrections represent spin/pseudospin-dependent scattering commonly termed skew scattering and side jump. They are associated with the Fermi surface, and occur because the scattering potential also contributes to band mixing and inter-band coherence. In this connection, an alternative interpretation is that disorder mixes the bands in the crystal, giving a correction to the wave function, and this results in a correction to the band-expectation values of certain physical observables, which is proportional to  the disorder strength. At the same time, an electric field shifts the Fermi surface away from equilibrium, and this shift is inversely proportional to the disorder strength. The net result in the average of physical observables is the product of the correction to the band-expectation value and the shift in the distribution function, and this product is formally independent of the disorder strength.


\begin{figure}[tbp!]
\begin{center}
\includegraphics[trim=0cm 0cm 0cm 0cm, clip, width=0.48\columnwidth]{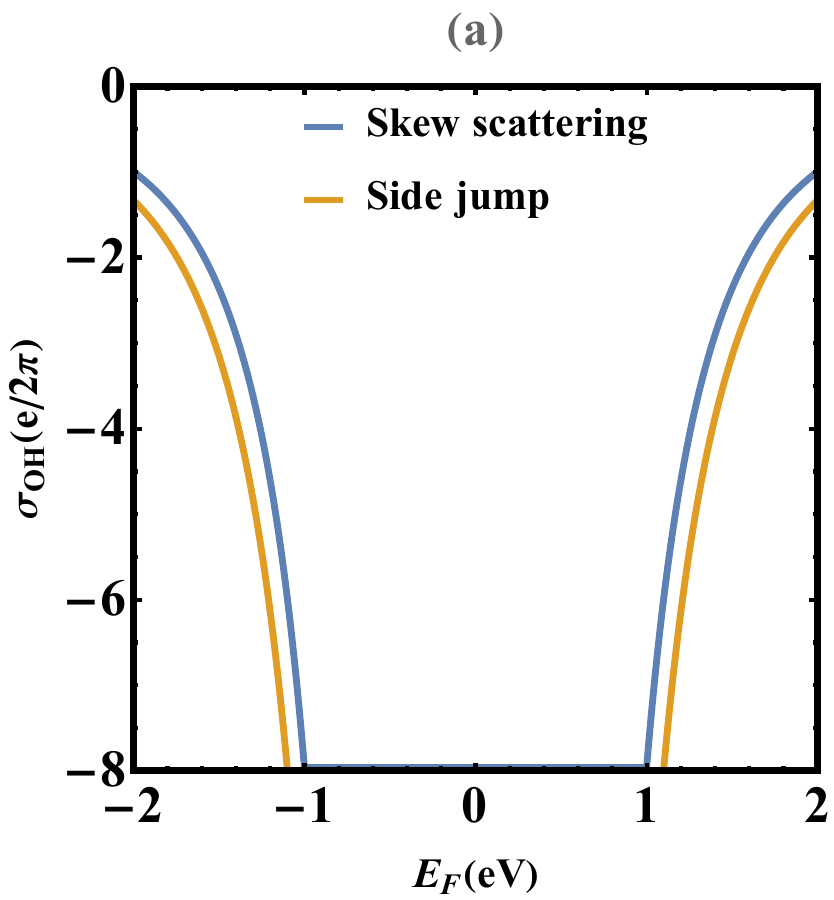}
\includegraphics[trim=0cm 0cm 0cm 0cm, clip, width=0.48\columnwidth]{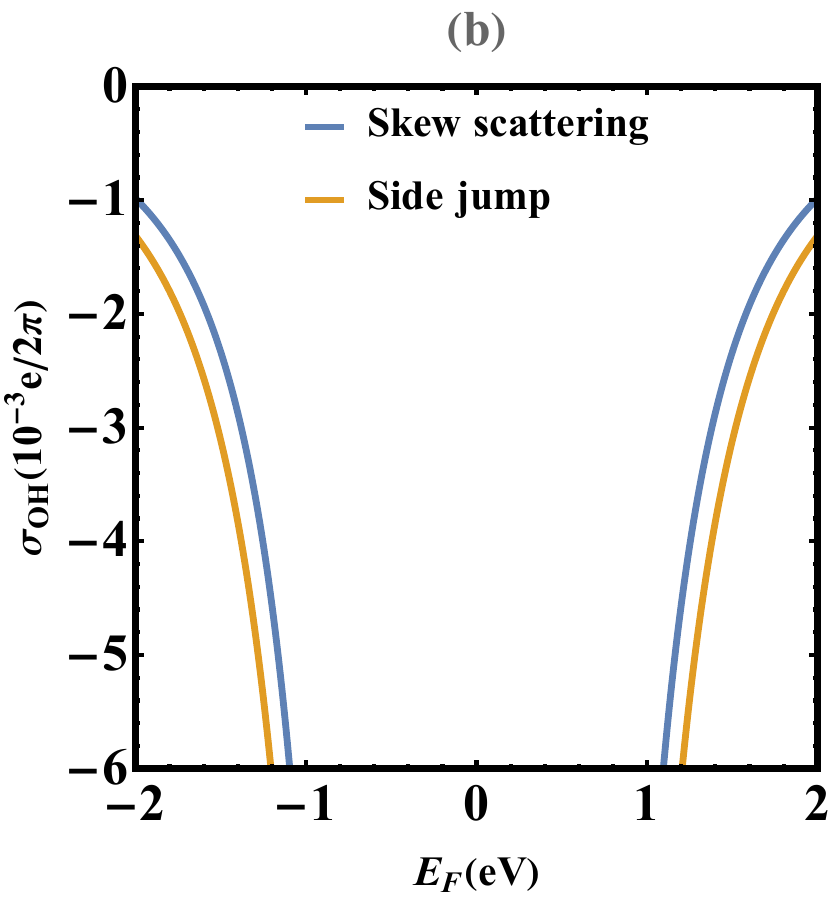}
\caption{\label{Fig3} The skew scattering and side jump contributions to OH-conductance $\sigma_\text{OH}$ vs the Fermi energy $E_F$: (a) for the 2H-phase monolayer MoS$_2$ with the parameters $v_F=3.6\text{eV}\AA$ and $\Delta=0.883\text{eV}$; (b) for the tilted 2D massive Dirac model with $v_F=1\text{eV}\AA$ and $\Delta=0.1\text{eV}$ and $v_t=0.1v_F$.}
\end{center}
\end{figure}
The intrinsic OHE arises from the intrinsic driving term $D^{mm'}_{E\bm{q}}=\frac{e \bm{E}}{\hbar} \cdot \big\{i\bm{\mathcal{R}}^{mm'}_{\bm q} [f^{(0)}(E^m_{\bm{q}})-f^{(0)}(E^{m'}_{\bm{q}})]\big\}$ with $f^{(0)}(E^m_{\bm{q}})$ the Fermi distribution function in equilibrium. The intrinsic OH-conductance is 
\begin{equation}
\ba
 \sigma^\text{int}_\text{OH}&\dps =-\frac{g_sg_\nu}{4\pi} \frac{em_e v^2_F}{3\hbar^2 \Delta} \frac{\Delta^3}{(v^2_Fq^2_F+\Delta^2)^{3/2}}.
  \ea
 \end{equation}
The side jump is further separated into two contributions: one is from anomalous driving term $D'_{E}$ and another is from $J_E(\langle\hat{\rho}\rangle)$ \cite{JE-PRR-Rhonald-2022}. The total side jump contribution is 
\begin{equation}\label{eq:SJ}
\ba
 \sigma^\text{sd}_\text{OH}&\dps=-\frac{g_sg_\nu}{4\pi} \frac{e m_e v^2_F}{3\hbar^2\Delta }\frac{12\Delta^3}{(v^2_Fq^2_F+\Delta^2)^{3/2}}.
 \ea
 \end{equation}
The skew scattering contribution stems from  $-J^\text{1st}_0 [S^{(0)}_{{E}}]$, 
 \begin{equation}\label{eq:SS}
\ba
 \sigma^\text{sk}_\text{OH}&\dps =-\frac{g_sg_\nu}{4\pi} \frac{em_e v^2_F}{3\hbar^2 \Delta} \frac{9\Delta^3}{(v^2_Fq^2_F+\Delta^2)^{3/2}}.
 \ea
 \end{equation}
 The total extrinsic OHE is $\sigma^\text{ext}_\text{OH}=\sigma^\text{sd}_\text{OH}+\sigma^\text{sk}_\text{OH}=21\sigma^\text{int}_\text{OH}$. The individual contributions are plotted in Fig.~\ref{Fig3}.

\textit{Topological antiferromagnets}. These are important for orbital torque applications: i. an external electric field can interact directly with orbital degrees of freedom without requiring the spin-orbit interaction, electrical generation of OAM can rely on light atomic elements \cite{ISHE-IOHE-PRB-2008-Inoue,OHE-PRL-2009-Inoue,OHE-metal-PRM-2022-Oppeneer,IOHE-PRB-2021-Amit}, 
widening the material choice for the electrical control of magnetism \cite{OT-FM-PRB-2021-YoshiChika, LS-conversion-CP-2021-Byong-Guk, Large-OAM-NL-2023-Piamonteze}. ii. the absence of spin-orbit coupling could result in long-lived magnetic information \cite{OAM-Exp-Tobias}, the orbital torque is investigated as an alternative to the spin torque, which has applications in data storage and nonvolatile logic \cite{CI-SOT-RMP-2019-Manchon,Roadmap-SOT-Review,SOT-topo-material-Rev}.  Topological antiferromagnets are described by a tilted Dirac Hamiltonian \cite{lahiri_PRB2021_nonlinear}:
\begin{equation}
 H_0({\bm q})= \nu v_t q_x \tau_0+ v_F( q_x \tau_x -\nu q_y\tau_y) +\Delta\tau_z,
\end{equation}
where $v_t$ introduces a tilt along $q_x$-axis. This Hamiltonian breaks both $\mathcal{T}$- and $\mathcal{P}$- symmetry. With $\nu=1$, the dispersion is $E^{\pm}_{\bm q}= v_tq_x\pm E_q$. The tilt will get into the energy conservation function in the scattering integral and Fermi distribution function, while the tilt doesn't get into the eigenstates. To determine the OHE for the tilted 2D massive Dirac system one can expand the tilt up to first order in the energy function by assuming $v_t/v_F\ll1$. So in the scattering process, the energy conservation function can be expanded as $\delta(E^m_{\bm q}-E^m_{{\bm q}'}) \approx \delta(E_q-E_{q'})+v_tq\big(\cos\phi-\cos\phi'\big)\pd{\delta(E_q-E_{q'})}{E_q}$. The Fermi distribution function in equilibrium and at the absolute zero of temperature can also be expanded $f^{(0)}(E^+_{\bm q}) \approx \Theta(E_F-E^+_{\bm q})=\Theta(E_F-E_{q})+v_tq\cos\phi \pd{\Theta(E_F-E_{q})}{E_q}$. Then the density matrix can be decomposed as $\rho_E+\rho_{E,v_t}$ with $\rho_{E,v_t}$ the correction to first order in the tilt. The expansion involves an angular factor, which, in the operator trace, is further multiplied by angular factors in the density matrix and velocity operator. The product of these three angular factors causes the correction to the OHE to first order in the tilt to vanish. Therefore, our calculation of the OHE in the 2D massive Dirac model applies to the tilted 2D massive Dirac model as well, up to terms of second order in the tilt. Based on the values in Refs.~\cite{XiaoDi-PhysRevLett.127.277201,npj-CuMnAs-vF}, we expect such terms to account for $<5\%$ of the total.


\textit{Experimental observation}. In systems with topological textures disorder yields contributions formally of zeroth order in the disorder strength, which are hard to distinguish experimentally from the intrinsic terms in the DC regime. 
However, it should be possible to distinguish the intrinsic and extrinsic contributions to the OHE using AC techniques. This would be the orbital transport analogue of the magneto-optical Kerr effect \cite{PhysRevLett.105.057401}, which was shown to be a strong probe of intrinsic transport in the anomalous Hall effect of topological insulators. Since disorder effects are suppressed as $1/(\omega\tau)$, with $\omega$ the light frequency, one requires $\omega\tau \gg 1$. Taking $\tau \sim 0.1\text{ps}$ places the wavelength in the ultraviolet range.

\textit{Comparison with previous work.} The total OH-conductance calculated here also reflects knowledge built up during earlier studies of the anomalous and spin-Hall effects \cite{Interband-Coherence-PRB-2017-Dimi, Inoue-SHE-PhysRevB.70.041303, Inoue-AHE-PhysRevLett.97.046604, AHE-Dirac-PRB-2007-Sinova}. In the spin Hall effect disorder corrections cancel the intrinsic contribution altogether, while in the anomalous Hall effect they overwhelm the intrinsic contribution by a factor of 7 \cite{AHE-Dirac-PRB-2007-Sinova,JE-PRR-Rhonald-2022}. Here the extrinsic OHE is 21 times the intrinsic contribution. Recalling that for Dirac fermions the velocity and spin operators coincide, these factors are understandable in light of the qualitative difference between the orbital and spin angular momentum operators \cite{ODynamics-PRL-2022-Kyoung-Whan}. The intrinsic and extrinsic terms discussed here form the main contributions to the OHE in the \textit{good metal} regime, in the sense used in Ref.~[\onlinecite{AHE-RevModPhys.82.1539}] for the anomalous Hall effect. Disorder effects beyond the first Born approximation, including correlations, may be important in certain regimes \cite{AHE-Beyond-Born-EPL-2015-Titov,EAHE-PRB-2023-ChenWei,AHE-Disorder-Corr-PRB-2017-Titov}. In particular, an additional skew scattering term appears in the second Born approximation, yielding a contribution to the OH-conductance $ \propto \frac{U^3_1}{2\pi n_i U^4_0}\frac{m_e v^2_F\Delta^2}{(q^2_Fv^2_F+\Delta^2)}$, which dominates as the ballistic limit is approached. The large extrinsic OHE in graphene and monolayer TMDs systems is due to their Dirac dispersion \cite{MoS2-Band-NL-2012} and inversion symmetry breaking.

The extrinsic OHE, including skew-scattering and side-jump, has been considered theoretically by Bernevig \textit{et al.} in hole-doped Si \cite{Orbitronics-PRL-2005-Shoucheng}. The effect of impurities in the Kubo linear response theory of Ref.~\cite{Orbitronics-PRL-2005-Shoucheng} is twofold: through the vertex correction to the current operator and the self-energy. Bernevig \textit{et al.} showed that vertex corrections from impurity scattering vanish for a low-energy model for hole-doped Si. However, a recent work \cite{Tangping} shows this is not the case in generic inversion symmetric systems.  

\textit{Application of the theory.} Some words are in order regarding the range of applicability of our theory. Firstly, since most of the orbital moment contribution comes from the valley points \cite{TMDsoc-IOHE-PRB-Satpathy-2020}, we considered the OHE from the $K$ and $K'$ points, without including the contribution from the remainder of the Brillouin zone. Secondly, we approximate the Fermi-Dirac function as a step function  which is the zero-temperature limit, whereas orbital Hall measurements are performed at room temperature i.e. 25 meV \cite{Exp-OHE-Ti-Nat-2023-Hyun-Woo,OOS-Cvert-2020-PRL-Mathias}. Our main assumption is that the fermion system is degenerate, i.e. $E_F\gg k_BT$. However, based on realistic values for $\tau$, our theory requires the Fermi energy to be away from the band edge by at least 50 meV in order to satisfy $E_F\tau/\hbar \gg 1$. This is twice room temperature: whereas the assumption of degeneracy is not perfect, the system is still approximately degenerate. Thirdly, we have assumed short-range disorder. For a more generic disorder model, although we do not expect the results to change qualitatively, the RHS of Eqs.~(\ref{eq:SJ}) and~(\ref{eq:SS}) will depend on the angular characteristics of the scattering potential. Finally but most importantly, the theory requires $E_F\tau/\hbar \gg 1$, which excludes the low-density regions in the vicinity of the conduction and valence band edges. These regions are virtually impossible to capture by most theoretical approaches. This is one of the fundamental paradoxes of transport theory: disorder is needed on physical grounds, but it needs to satisfy $E_F\tau \gg 1$ (alternatively $q_F l \gg 1$, where $l$ is the mean free path) for the theory to be applicable. The appearance of a term $\propto \tau$ (i.e. $\propto 1/n_i$) is understood on physical grounds: it represents the shift in the Fermi surface induced by the electric field. The fact that this shift is $\propto \tau$ reflects the need for disorder to be present so as to keep the Fermi surface near equilibrium. If disorder is removed entirely and $\tau$ formally tends to infinity this would imply that an infinitesimally small electric field can shift the Fermi surface infinitely far away from equilibrium. Such a process is unphysical, therefore the limit $\tau\rightarrow \infty$ is not captured in this formulation of transport theory. The resolution is that the size of the sample now needs to be considered and it is assumed that the mean free path is larger than the size of the sample, so the system transitions into the Landauer-Buttiker regime. This regime can also be captured in our density matrix language \cite{PhysRevLett.127.206801}, but is unlikely to be relevant to the materials studied in the present work, where transport is diffusive. The opposite limit, $q_F \rightarrow 0$, is equivalent to $E_F \tau \ll 1$, which represents the transition to the localised regime. The transport theory first needs to be augmented with the Cooperon in order to describe weak localisation \cite{Liu_2023}.

\textit{Conclusions}. We have determined the disorder contribution to the OHE of 2D massive Dirac fermions and shown that it exceeds the intrinsic contribution by an order of magnitude, using graphene, transition metal dichalcogenides and topological antiferromagnets as prototypes. The calculation offers an approach to tuning and maximizing the orbital torque and can be extended to other classes of materials, such as Weyl and Dirac semimetals, topological insulators, and van der Waals heterostructures, as well as opening future perspectives on graphene-based orbitronics and twistronics \cite{Twistnoics-Allan}. 

\textit{Acknowledgments.} This work is supported by the Australian Research Council Centre of Excellence in Future Low-Energy Electronics Technologies, project number CE170100039. We acknowledge enlightening discussions with Di Xiao, Giovanni Vignale, Eli Zeldov, Binghai Yan, Amit Agarwal and Kamal Das. 


%

\end{document}